\documentclass[twocolumn,nofootinbib,showpacs,amsmath,amssymb,aps,prd,blancelastpage,superscriptaddress]{revtex4-1}

\usepackage{amssymb} \usepackage{amsmath} \usepackage{graphicx}
\usepackage{epsfig,latexsym}

\RequirePackage{xspace} \allowdisplaybreaks

\begin{document}

\def\bef{\begin{figure}}
\def\eef{\end{figure}}

\newcommand{\nl}{\nonumber\\}
\newcommand{\ans}{ansatz }
\newcommand{\be}[1]{\begin{equation}\label{#1}}
\newcommand{\beq}{\begin{equation}}
\newcommand{\ee}{\end{equation}}
\newcommand{\beqn}[1]{\begin{eqnarray}\label{#1}}
\newcommand{\eeqn}{\end{eqnarray}}
\newcommand{\bd}{\begin{displaymath}}
\newcommand{\ed}{\end{displaymath}}
\newcommand{\mat}[4]{\left(\begin{array}{cc}{#1}&{#2}\\{#3}&{#4}
\end{array}\right)}
\newcommand{\matr}[9]{\left(\begin{array}{ccc}{#1}&{#2}&{#3}\\
{#4}&{#5}&{#6}\\{#7}&{#8}&{#9}\end{array}\right)}
\newcommand{\matrr}[6]{\left(\begin{array}{cc}{#1}&{#2}\\
{#3}&{#4}\\{#5}&{#6}\end{array}\right)}
\newcommand{\cvb}[3]{#1^{#2}_{#3}}
\def\lsim{\raise0.3ex\hbox{$\;<$\kern-0.75em\raise-1.1ex
e\hbox{$\sim\;$}}}
\def\gsim{\raise0.3ex\hbox{$\;>$\kern-0.75em\raise-1.1ex
\hbox{$\sim\;$}}}
\def\abs#1{\left| #1\right|}
\def\simlt{\mathrel{\lower2.5pt\vbox{\lineskip=0pt\baselineskip=0pt
           \hbox{$<$}\hbox{$\sim$}}}}
\def\simgt{\mathrel{\lower2.5pt\vbox{\lineskip=0pt\baselineskip=0pt
           \hbox{$>$}\hbox{$\sim$}}}}
\def\unity{{\hbox{1\kern-.8mm l}}}
\newcommand{\eps}{\varepsilon}
\def\ep{\epsilon}
\def\ga{\gamma}
\def\Ga{\Gamma}
\def\om{\omega}
\def\omp{{\omega^\prime}}
\def\Om{\Omega}
\def\la{\lambda}
\def\La{\Lambda}
\def\al{\alpha}
\newcommand{\ov}{\overline}
\renewcommand{\to}{\rightarrow}
\renewcommand{\vec}[1]{\mathbf{#1}}
\newcommand{\vect}[1]{\mbox{\boldmath$#1$}}
\def\tm{{\widetilde{m}}}
\def\mcirc{{\stackrel{o}{m}}}
\newcommand{\Dm}{\Delta m}
\newcommand{\dm}{\varepsilon}
\newcommand{\tanb}{\tan\beta}
\newcommand{\nbar}{\tilde{n}}
\newcommand\PM[1]{\begin{pmatrix}#1\end{pmatrix}}
\newcommand{\up}{\uparrow}
\newcommand{\down}{\downarrow}
\def\omE{\omega_{\rm Ter}}

\newcommand{\Dsusy}{{susy \hspace{-9.4pt} \slash}\;}
\newcommand{\DCP}{{CP \hspace{-7.4pt} \slash}\;}
\newcommand{\mc}{\mathcal}
\newcommand{\gr}{\mathbf}
\renewcommand{\to}{\rightarrow}
\newcommand{\gtc}{\mathfrak}
\newcommand{\wh}{\widehat}
\newcommand{\br}{\langle}
\newcommand{\kt}{\rangle}

\def\lsim{\mathrel{\mathop  {\hbox{\lower0.5ex\hbox{$\sim$}
\kern-0.8em\lower-0.7ex\hbox{$<$}}}}}
\def\gsim{\mathrel{\mathop  {\hbox{\lower0.5ex\hbox{$\sim$}
\kern-0.8em\lower-0.7ex\hbox{$>$}}}}}

\def\nn{\\  \nonumber}
\def\de{\partial}
\def\brf{{\mathbf f}}
\def\bbf{\bar{\bf f}}
\def\bF{{\bf F}}
\def\bbF{\bar{\bf F}}
\def\bA{{\mathbf A}}
\def\bB{{\mathbf B}}
\def\bG{{\mathbf G}}
\def\bI{{\mathbf I}}
\def\bM{{\mathbf M}}
\def\bY{{\mathbf Y}}
\def\bX{{\mathbf X}}
\def\bS{{\mathbf S}}
\def\bb{{\mathbf b}}
\def\bh{{\mathbf h}}
\def\bg{{\mathbf g}}
\def\bla{{\mathbf \la}}
\def\bmu{\mathbf m }
\def\by{{\mathbf y}}
\def\bmu{\mbox{\boldmath $\mu$} }
\def\bsig{\mbox{\boldmath $\sigma$} }
\def\bunity{{\mathbf 1}}
\def\cA{{\cal A}}
\def\cB{{\cal B}}
\def\cC{{\cal C}}
\def\cD{{\cal D}}
\def\cF{{\cal F}}
\def\cG{{\cal G}}
\def\cH{{\cal H}}
\def\cI{{\cal I}}
\def\cL{{\cal L}}
\def\cN{{\cal N}}
\def\cM{{\cal M}}
\def\cO{{\cal O}}
\def\cR{{\cal R}}
\def\cS{{\cal S}}
\def\cT{{\cal T}}
\def\eV{{\rm eV}}
%

\title{Probing Trans-electroweak First Order Phase Transitions from Gravitational Waves}

\author{Andrea Addazi}
\email{andrea.addazi@lngs.infn.it}
\affiliation{Department of Physics \& Center for Field Theory and Particle Physics, 
Fudan University, 200433 Shanghai, China}

\author{Antonino Marcian\`o}
\email{marciano@fudan.edu.cn}
\affiliation{Department of Physics \& Center for Field Theory and Particle Physics, 
Fudan University, 200433 Shanghai, China}

\author{Roman Pasechnik}
\email{Roman.Pasechnik@thep.lu.se}
\affiliation{Department of Astronomy and Theoretical Physics, Lund
University, SE-223 62 Lund, Sweden}
\affiliation{Nuclear Physics Institute ASCR, 25068 \v{R}e\v{z}, 
Czech Republic}

\begin{abstract}
\noindent
We propose direct tests of very high energy first-order phase transitions, which are elusive to collider physics, deploying the gravitational waves measurements. We show that first-order phase transitions lying into a large window of critical temperatures, which is considerably larger than the electroweak energy scale, can be tested from aLIGO and Einstein Telescope. This provides the possibility to probe several inflationary mechanisms ending with the inflaton in a false minimum, and high-energy first order phase transitions that are due to new scalar bosons, beyond the Standard Model of particle physics. As an important example, we consider the axion monodromy inflationary scenario, and analyze the potential for its experimental verification, deploying the gravitational wave interferometers. 
\end{abstract}

\maketitle

{\it Introduction.---} The first direct detections of gravitational waves (GW) from a merging of black holes and neutron stars, as recently measured by the LIGO/VIRGO collaboration, raise an urgent question: can we detect any hint of new physics beyond the Standard Model (SM) of particles and cosmology within the same frequency sensitivity range of LIGO/VIRGO?

In this letter, we show that first order phase transitions (FOPT) at very high energy, beyond the electroweak scale, can be tested at the next generation of experiments beyond LIGO and VIRGO, including the aLIGO and Einstein Telescope (ET). According to the standard scenario of cosmological FOPTs, our Universe was initially set in a false vacuum state. After a certain time scale, a tunneling toward the true vacuum state happened. This induced the nucleation of bubbles, relativistically expanding with constant acceleration. The violent acceleration 
was sourced by a difference of pressure between the exterior false vacuum and the interior true vacuum. Such bubbles nucleating in the early Universe have produced the stochastic GW background through three distinct processes: bubble-bubble scattering, acoustic shock waves and magnetohydrodynamic (MHD) turbulence.

From the experimental perspective, the temperature of the FOPT determines the characteristic frequency window of GWs. For instance, the phase transitions with critical temperatures around $100\, {\rm GeV} \div 1{\rm TeV}$ produce the GW signals that are peaked within the frequency range of $1 \div 10\, {\rm mHz}$
\cite{Witten:1984rs, Turner:1990rc, Hogan:1986qda, Kosowsky:1991ua, Kamionkowski:1993fg, Hindmarsh:2013xza, Hindmarsh:2015qta, Delaunay:2007wb}. Intriguingly, these frequencies can be probed by the next generation of interferometers, including LISA, BBO and U-DECIGO \cite{Caprini:2015zlo, Kudoh:2005as, Audley:2017drz}\footnote{See also a more recent analysis 
in Refs.~\cite{Vieu:2018nfq,Vieu:2018zze,Mazumdar:2018dfl,Kobakhidze:2017mru,Gould:2019qek,Brdar:2018num} and Ref.\cite{Weir:2017wfa} for a complete review on recent developments in the subject. }. Several theoretical models associated with a FOPT at the energy scale of about $100\, {\rm GeV} \div 1{\rm TeV}$ were hitherto studied \cite{Caprini:2015zlo, Kudoh:2005as, Audley:2017drz}. Nonetheless, FOPTs with a critical temperature much larger than the electroweak scale cannot be observed by LISA, U-DECIGO and BBO. This seems to preclude a possibility of testing the very high-energy FOPTs due to new physics beyond the SM. 

In this work, we show that FOPTs at energy scales much higher than the electroweak energy scale lie in the frequencies' window of LIGO/VIRGO, KAGRA, aLIGO and ET. The FOPT scenarios can be very elusive while the current data from LIGO and VIRGO are considered. Specifically, aLIGO and ET may achieve very interesting sensitivity ranges, allowing to test FOPTs at energies much higher than the electroweak scale. 

Although these considerations may have very strong implications for probing the physics of early Universe, these were never discussed in the literature. By testing the cosmological FOPTs, we can infer precious information on such stages of the Universe evolution as Inflation, Grand Unified Theories (GUTs), Baryogenesis, Primordial Black Holes (PBHs) production etc. For instance, many models of Inflation, with the inflationary regime ending in a false minimum after reheating, can be tested and eventually falsified \cite{Inflation1,Inflation2,Turner:1992tz,Lopez:2013mqa}. Several GUT models, including the $SO(10)$ Pati-Salam, predict a high energy pattern of FOPTs sourced from a rich Higgs sector --- for example, as the spontaneous symmetry breaking pattern $SO(10)\rightarrow $ $SU(4)\times SU(2)_{L}\times SU(2)_{R} \rightarrow $ $SU_{c}(3)\times U_{B-L}(1)\times SU(2)_{L}\times SU(2)_{R} \rightarrow $ $SU_{c}(3) \times SU_{L}(2)\times U_{Y}(1)$ \cite{PS} realizes. On the other hand, PBHs may compose a part of dark matter, and be efficiently produced from FOPTs, through bubble-bubble scatterings \cite{Khlopov:1999ys}. 

The large variety of possible instantiations of cosmological FOPTs at energies beyond the electroweak scale calls for a model independent analysis, in the framework of effective field theories. We  investigate a space of parameters that corresponds to a large class of models related to high-energy FOPTs. As an inverse scattering approach, our analysis can be useful in order to infer new bounds on the inflaton potentials, as well as to constraint extensions of the Higgs sector beyond the SM, deploying the GWs data from LIGO and VIRGO, and future data from other experiments, probing the same frequency window --- but with more sensitivity in the GWs energy-density. Elusive to any possible laboratory experiment, this possibility is certainly intriguing and worth a thorough discussion.

\vspace{0.1cm}

{\it GW signals from FOPT.---}  The FOPTs in the early Universe can produce vacuum bubbles that expand at relativistic velocities. Such a mechanism produces a stochastic background of gravitational radiation. GWs are sourced via bubble-bubble collisions, sound waves and turbulence generated by the bubbles' expansion in the plasma. 
\begin{figure}[t]
\centerline{ \includegraphics [height=4.5 cm,width=0.96\columnwidth]{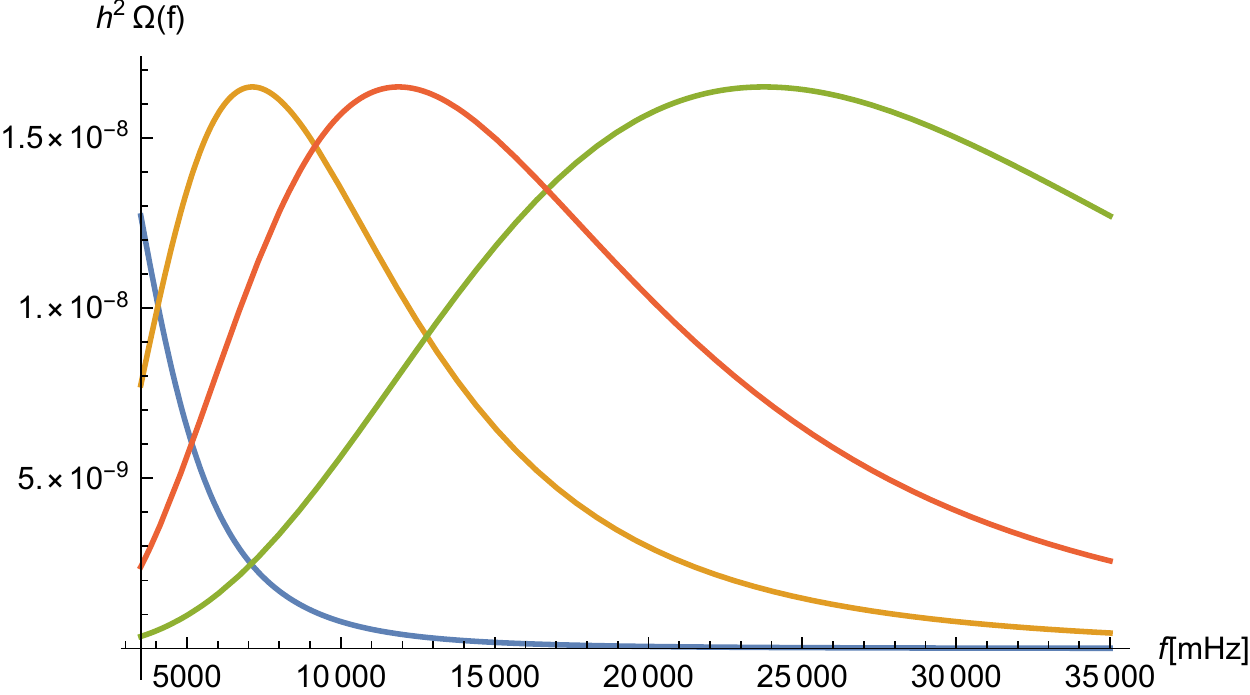}}
\caption{Examples of non-runaway cases are displayed, with the same value of the parameters $v_{w}=0.8$, $\alpha=0.9$, $g_{*}\simeq g_{SM}$ and $\beta/H_{*}=10$, but with varying FOPT temperature, namely $T_{*}/(10^{8}\, {\rm GeV})=\{0.1, 0.3, 0.5, 5\}$, corresponding to the blue, orange, red and green lines, respectively. }
\label{plot}
\end{figure}
\begin{figure}[t]
\centerline{ \includegraphics [height=4.5 cm,width=0.96\columnwidth]{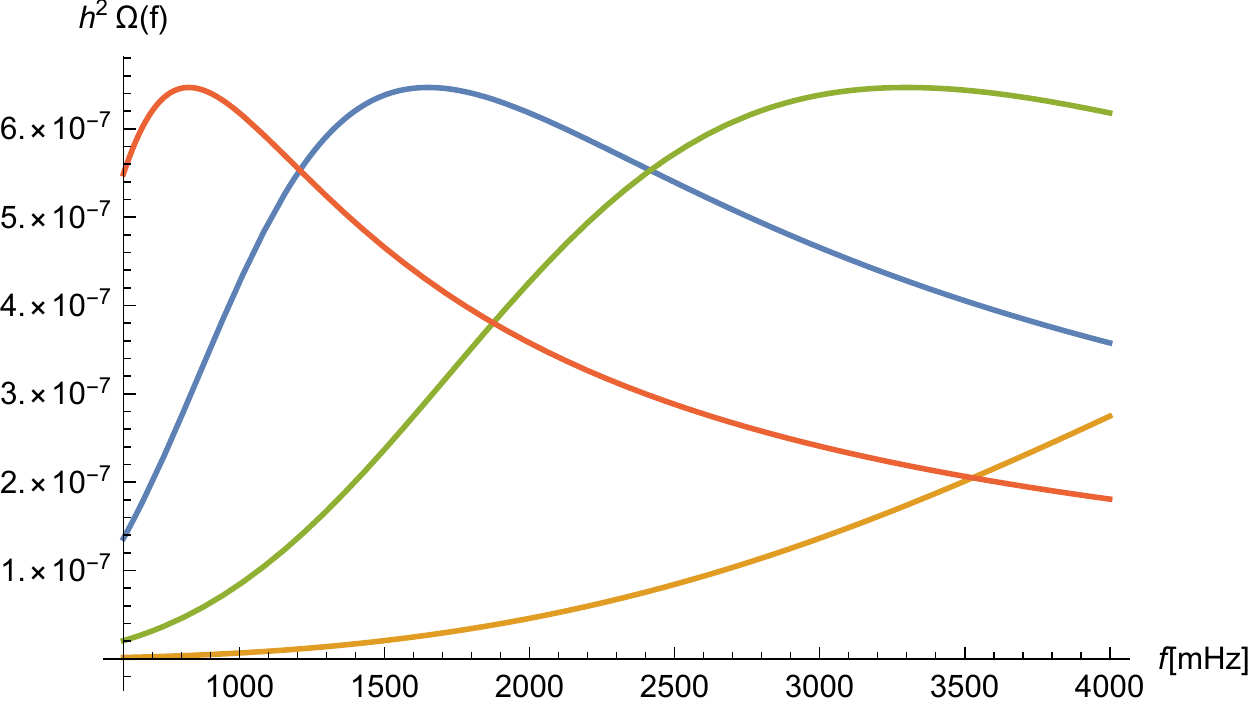}}
\caption{Examples of runaway cases are displayed, with same $v_{w}=1$, $\alpha=1$, $g_{*}\simeq g_{\rm SM}$, $\beta/H_{*}=10$ and $T_{*}/(10^{8}\, {\rm GeV})=\{0.5,1,2,5\}$ in red, blue, green, orange lines, respectively. }
\label{plot}
\end{figure}
The three contributions to the GW spectrum are related to the effective potential of the scalar field, $V_{\rm eff}(T)$, which is responsible for the bubbles coalescence. In the early Universe plasma, the scalar potential receives extra contributions that depend on the plasma temperature and the coupling constants to the other fields. 

The bubble nucleation has a rate that formally reads as $\Gamma(t)=A(t)e^{-S(t)}$, where $S$ is the Euclidean action describing the bubble dynamics. From this expression, one can define a critical temperature $T_{*}$ for the FOPT, at which the nucleation rate becomes large enough to induce at least the nucleation of one bubble per horizon volume with a probability of order $1$. From the $\Gamma(t)$ rate, one therefore defines   
\begin{eqnarray}
\label{beta}
\beta=-\left[\frac{dS}{dt}\right]_{t=t_{*}}\simeq \frac{\dot{\Gamma}}{\Gamma}\, , 
\end{eqnarray}
as an inverse characteristic time-scale of the FOPT. The hierarchy among the $\beta$ and the Hubble rate at the FOPT temperature $H_{*}\equiv H(T_{*})$ is crucial to ensure an efficient production of GWs. The ratio can be cast as $$\frac{\beta}{H_{*}}=T_{*}\left[\frac{dS}{dT}\right]_{T=T_{*}}\, .$$

Another important parameter is 
\begin{eqnarray}
\label{alpha}
\alpha=\frac{\rho_{\rm vac}}{\rho_{\rm rad}(T_{*})}\, , 
\end{eqnarray}
which measures the vacuum energy density released in the FOPT compared to the energy density of radiation at $T_{*}$  --- the latter is $\rho_{\rm rad}(T_{*})=g_{*}\pi^{2}T_{*}^{4}/30$, 
with $g_{*}$ denoting the number of cosmological relativistic degrees of freedom at $T_{*}$. Other useful parameters characterizing the GW spectrum are 
\begin{eqnarray}
\label{alpha}
\kappa_{\phi}=\frac{\rho_{\phi}}{\rho_{\rm vac}},\,\,\,\, \qquad \kappa_{v}=\frac{\rho_{v}}{\rho_{\rm vac}}\, , 
\end{eqnarray}
measuring the inverse fractions of the vacuum energy converted into gradient energy of the scalar field and into the fluid motion, respectively. Finally, a relevant parameter characterizing the GW spectrum is the bubble wall velocity, $v_w$, in the rest frame of the fluid that is found asymptotically far from the bubble. 

It is worth to note that all these quantities have direct relations to the thermally corrected scalar potential, appearing in the Euclidean action and in the released vacuum energy density, i.e. 
\begin{align}
&\rho_{\rm vac}(T_{*})=- \Big[ T\frac{dV_{\rm eff}}{dT}-V_{\rm eff}(T) \Big]|_{T=T_{*}} ~, \nonumber \\
 &S(T) \simeq \frac{S_{3}(T)}{T} ~,~~ S_{3} \equiv \int d^{3}r [ \partial_{i}\sigma^{\dagger}\partial_{i}\sigma+V_{\rm eff}(\sigma,T) ] ~, \nonumber \\
 &\Gamma_{0}(T) \sim T^{4} ~,~~ \Gamma = \Gamma_{0}(T) \exp[-S(T)] \,, \nonumber
\end{align}
where $\sigma$ is the scalar field undergoing the FOPT. 

\vspace{0.1cm} 
 
{\it Bubble-bubble collision.---} The GW energy spectrum produced during the collision of two bubbles depends on the false-vacuum energy and the bubble-size at collision. This implies that the GW spectrum depends only on the ratio between the rate of bubbles nucleation and the Hubble rate of the Universe, at temperature of the FOPT, and on the latent heat energy and the bubble wall velocity. The redshift of GWs introduces a dependence on the number of cosmological degrees of freedom. During the FOPT, the GW intensity peak produced by the bubbles' collisions reads \cite{Kosowsky:1991ua,Kamionkowski:1993fg} 
\begin{eqnarray}
\label{C}
\frac{f_{*}}{\beta}=\left(  \frac{0.62}{1.8-0.1v_{w}+v_{w}^{2}}\right)\,,
\end{eqnarray}
where $f_{*}$ is determined by the characteristic time-scale of the FOPT, i.e.~$1/\beta$. Using the inverse Hubble time at the GW production, redshifted today, from the well known relation $$h_{*}\!=\!16.5 \times 10^{-3}\, {\rm mHz} \left(\frac{f_{*}}{\beta} \right)\!\left(\frac{\beta}{H_{*}}\right)\! \left( \frac{T_{*}}
{100\, {\rm GeV}}\right)\!\left(\frac{g_{*}}{100} \right)^{\frac{1}{6}}, $$
one obtains the frequency peak
\begin{eqnarray}
\label{CCC}
&&f_{\rm coll}=16.5 \times 10^{-3}{\rm mHz} \left( \frac{T_{*}}{100\, {\rm GeV}}\right) \left(\frac{g_{*}}{100} \right)^{\frac{1}{6}}\, .
\end{eqnarray}
This estimate was obtained assuming that the FOPT has happened in a radiation dominated epoch. However, in some particular instantiations beyond the SM this assumption may be relaxed. For a number of degrees of freedom that is close to the SM one, the GWs frequency ranges in the domain $1 \div 100\, {\rm Hz}$, which can be achieved for a range of temperatures $T_{*}\simeq 10^{7}\div 10^{9}\, {\rm GeV}$. This motivates our proposal in terms of order of magnitude.

At the same time, one should achieve an energy-density of GWs that is high enough to reach the sensitivity curves of the next generation of experiments probing the $1\div 100\, {\rm Hz}$ frequency range. The GW energy-density peak corresponds to \cite{Kosowsky:1991ua,Kamionkowski:1993fg}
\begin{eqnarray}
h^{2} \Omega_{\rm coll}(f) &=& 1.67 \times 10^{-5}\left( \frac{H_{*}}{\beta}\right)^{2}\left(\frac{\kappa \alpha}{1+\alpha} \right)^{2} \nonumber \\
&\times& \left( \frac{100}{g_{*}}\right)^{\frac{1}{3}} \left( \frac{0.11 v_{w}^{3}}{0.42+v_{w}^{2}}\right)\,,
\label{CCCC}
\end{eqnarray}
where $\kappa$ is the fraction of the latent heat deposited, in the envelope approximation, on the front of the FOPT. Assuming $\kappa \sim 10\%$, $H_{*}/\beta=1/100$, $v_{w}\sim 1$ and $g_{*}\sim g_{*}^{SM}\sim 100$, we can immediately estimate that $h^{2}\Omega_{coll}(f)$ reaches the range $10^{-9}\div 10^{-10}$. Although very far from the LIGO/VIRGO current sensitivity, this signal can be reached by the ET. More favorable cases are the ones with $\beta/H_{*}=1\div 10$, for which $h^{2}\Omega_{coll}(f)$ falls in the range $10^{-5}\div 10^{-7}$ --- all the other parameters being taken as above --- and these can be then tested by aLIGO.

\vspace{0.1cm}

{\it Sound waves.---} The estimates of the GW peak from sound waves is affected by $O(1)$ numerical uncertainties \cite{Cai:2017cbj,Ellis:2018mja}. The order of magnitude can be anyway inferred, considering the characteristic scale set by the average bubble separation, i.e. $R_{*}=(8\pi)^{1/3}v_{w}/\beta$. From a conservative estimate, the frequency peak reads 
$$f_{\rm sw}=\frac{2}{\sqrt{3}}\frac{\beta}{v_{w}}\,,$$ 
which, redshifted until the Universe today, reads 
\begin{eqnarray}
\label{SW}
f_{\rm sw}\!=\!1.9 \!\times \! 10^{-2} {\rm mHz}  \frac{1}{v_{w}} \!\! \left( \frac{\beta}{H_{*}}\right) \!\!\left( \frac{T_{*}}{100\, {\rm GeV}}\right) \!\! \left(\frac{g_{*}}{100} \right)^{\frac{1}{6}}\!.
\end{eqnarray}

Contrary to Eq.~(\ref{CCC}), in the latter expression there is also a dependence of the frequency peak on $\beta/H_{*}$. In the reasonable range $\beta/H_{*}=10 \div 1000$, the possible frequencies are $1\div 100\, {\rm Hz}$. For $v_{w}\simeq 1$, $\kappa_{v}\sim 1\div 10\%$, this latter corresponds to the range of temperatures 
$T_{*}=10^{6}\div 10^{10}\, {\rm GeV}$. Nonetheless, one should also consider that the energy-density scales are the inverse of $\beta/H_{*}$. In particular, the peak 
can be estimated to be 
\begin{eqnarray}
h^{2}\Omega_{\rm sw}=2.65 \times 10^{-6}\left( \frac{H_{*}}{\beta}\right) \nonumber \\
\times \left( \frac{\kappa_{v}\alpha}{1+\alpha}\right)^{2} \left(\frac{100}{g_{*}} \right)^{\frac{1}{3}}v_{w}\,.
\label{SW}
\end{eqnarray}

Detection from LIGO/VIRGO seems also in this case impossible, and very elusive even for aLIGO. Nonetheless, the ET can test this scenario for $\kappa_{v}\sim 10\%$, 
$v_{w}\sim 1$, $H_{*}/\beta\sim 1\div 10$, $T_{*}\sim 10^{7}\div 10^{8}\, {\rm GeV}$ and $\alpha \sim 1$, since the corresponding amplitudes would peak in 
the range $h^{2}\Omega_{sw}\sim 10^{-7}\div 10^{-8}$. 

\vspace{0.1cm}

{\it MHD turbulence.---} Finally, the frequency peak for MHD turbulence was recovered to be \cite{Caprini:2015zlo}
\begin{eqnarray}
f_{\rm turb}=2.7\times 10^{-2}\, {\rm mHz} \frac{1}{v_{w}}\! \left( \frac{\beta}{H_{*}}\right)\!\left(\frac{T_{*}}{100\,{\rm GeV}} \right)
\!\left(\frac{g_{*}}{100} \right)^{\frac{1}{6}} \,. \nonumber
\end{eqnarray}
In the reasonable domain $\beta/H_{*}=10 \div 1000$, one obtains possible frequencies in the $1\div 100\, {\rm Hz}$ range. Specifically, for $v_{w}\simeq 1$, the temperatures' 
range spans the region $T_{*}=10^{4}\div 10^{8}\, {\rm GeV}$. The GW energy-density peak is expressed by
 \begin{eqnarray}
\label{t1}
h^{2}\Omega_{\rm turb}\!=\!3.35 \times 10^{-4}\!\left( \frac{H_{*}}{\beta}\right)\!\left( \frac{\kappa_{t}\alpha}{1+\alpha}\right)^{\frac{3}{2}}\!v_{w}\!\left(\frac{100}{g_{*}} \right)^{\frac{1}{3}}\!\!,
\end{eqnarray}
where $\kappa_{t}$ denotes the fraction of latent heat transformed into turbulence of the plasma. Interesting situations similar to the ones already discussed 
for the case of sound waves can be recovered also for turbulence. Of course, approximations $O(1)$ are implied as well while considering the turbulence effects, 
and the competition of turbulence and sound waves is very much sensitive to the initial conditions. In both cases, these are issues not relevant to our discussion, 
since one cannot claim these predictions to be in the domain of high precision physics. 

\vspace{0.1cm}

{\it Dynamics of bubbles.---} The GW signals from FOPTs can radically differ from one another, depending on different regimes considered, in which either bubbles' collision, or MHD turbulence, or sound waves contributions may dominate. The dynamical regime is selected by the model-dependent component of the effective parameters introduced above. In our discussion, which is model independent, we will consider the most likely scenarios. These can be divided into two classes: i) non-runaway (in the plasma) bubbles; ii) runaway bubbles. 

\vspace{0.1cm}

{\it i) Non-runaway bubbles in the plasma.---} Bubbles expanding in a cosmic plasma can reach a relativistic velocity. In this situation, the energy stored in the scalar field can be neglected, since it would only be scaled with the surface of the bubble despite of its volume. The most relevant contributions to the signal are expected from the cosmic fluid motion induced from the bubbles' expansion, {\it i.e.} from sound waves and/or MHD turbulence. The energy-density spectrum can be approximated to be
\begin{equation}
\label{energydensity}
h^{2}\Omega_{GW}\simeq h^{2}\Omega_{turb}+ h^{2}\Omega_{sw}\, . 
\end{equation}
For this estimate, it is relevant to include $\kappa_{v}$ as the efficiency factor of conversion of the latent heat into the bubbles' motion through the bulk. The efficiency factor of non-runaway bubbles can be modeled with a good approximation, and is found to be in the two opposite velocity regimes
\begin{eqnarray}
&&\kappa_{v}\simeq \alpha (0.73+0.083\sqrt{\alpha}+\alpha)^{-1},\,\,\,\,\,\quad  v_{w}\sim 1\, , \nonumber \\
&&
\kappa_{v}\simeq v_{w}^{6/5}6.9 \alpha (1.36-0.037\sqrt{\alpha}+\alpha)^{-1},\,\,\,\,\quad  v_{w} \leq 1\,.  \nonumber
\end{eqnarray}
The efficiency of the energy transfer to the turbulence is then $\kappa_{t}\simeq \epsilon \kappa_{v}$, such that $\epsilon$ denotes the efficiency for the transfer of the bulk motion into the plasma turbulence --- from numerical simulations, one finds that $\epsilon \sim 5\div 10 \%$ \cite{Caprini:2015zlo,Espinosa:2010hh}. In Fig.~1, we display several simulations derived within this regime. We show that in some viable cases, sensitivity curves of the ET can be reached. 

\vspace{0.1cm}

{\it ii) Runaway bubbles in the plasma.---} Another possibility is that the bubble wall experiences unbound acceleration, running away with $v_{w}\rightarrow 1$. This case corresponds to a very high energy density stored in the scalar field profile, {\it i.e.} for $\alpha\gg1$
\begin{eqnarray}
h^{2}\Omega_{\rm GW}\simeq h^{2}\Omega_{\phi}+h^{2}\Omega_{\rm sw}+h^{2}\Omega_{\rm turb}\,, \nonumber
\end{eqnarray}
where $\Omega_{\phi}$ is sourced by the gradients of the scalar field, modeled in the envelope approximation. The variable $\alpha_{\infty}$ is defined as the minimum threshold value of $\alpha$ such that the bubbles start to run away. 
The efficiency of the energy transfer is parametrized by 
\begin{eqnarray} 
&&\kappa_{\phi}=(\alpha-\alpha_{\infty})/\alpha\geq 0,\,\,\,\,\qquad \kappa_{v}=\frac{\alpha_{\infty}}{\alpha}\kappa_{\infty}\, ,\nonumber\\
&&\kappa_{t}=(1-\alpha_{\infty})\frac{\alpha_{\infty}}{\alpha},\,\,\,\,\,\,\kappa_{\infty}=\frac{\alpha_{\infty}}{0.73+0.083\sqrt{\alpha_{\infty}}+\alpha_{\infty}}\, . \nonumber
\end{eqnarray} 
 The $\alpha_{\infty}$ parameter is model dependent, and reads
 \begin{equation}
\label{kappav}
 \alpha_{\infty}\simeq \frac{30}{24\pi^{2}}\frac{\sum_{i}c_{i}\Delta m_{i}^{2}(\phi_{*})}{g_{*}T_{*}^{2}}\, , 
 \end{equation}
where $\phi_{*}$ is the field value acquired inside the bubble immediately after the tunneling process, $i$ runs all over the particles, $\Delta m_{i}^{2}(\phi_{*})$ are the squared mass differences in the two phases, $c_{i}$ is equal to $N_{i}$ for bosons and $N_{i}/2$ for fermions in terms of the numbers of species $N_{i}$. In Fig.~2, we display several cases of runaway bubbles. 

\vspace{0.1cm}

{\it Trans-electroweak phase transitions in axion monodromy inflation.---} We now focus on the well known case of axion monodromy inflation, where the inflaton has a potential of the form 
\begin{equation}
\label{potential}
V(a)=\frac{1}{2}m^{2}a^{2}+\Lambda^{4}\cos\Big(\frac{a}{f}+\bar{\phi} \Big)\, ,
\end{equation}
$f$ and $m$ denoting the axion decay constant and mass, $\Lambda$ standing for the typical non-perturbative axion scale --- this analogous to the QCD dimensional 
transmutation energy scale of the QCD axion. 
\begin{figure}[!h]
\begin{center}
\includegraphics[width=0.8\columnwidth]{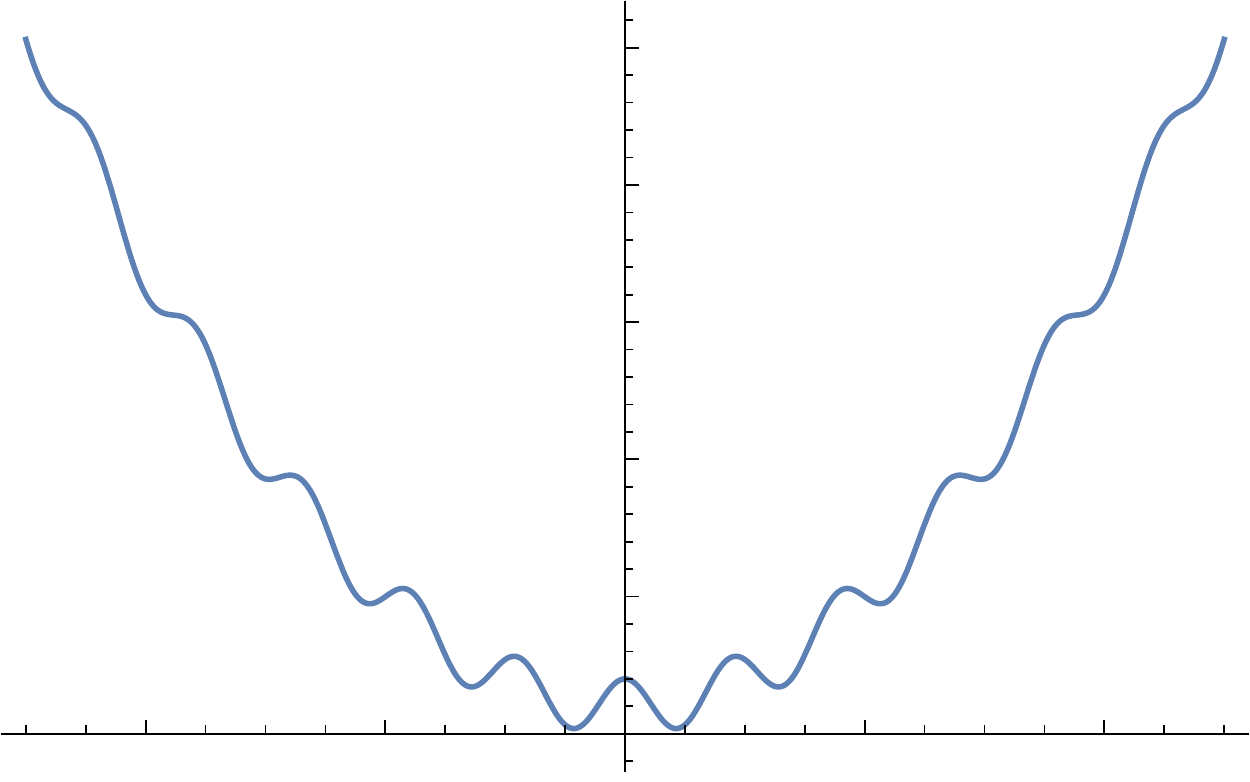} 
\includegraphics[width=0.8\columnwidth]{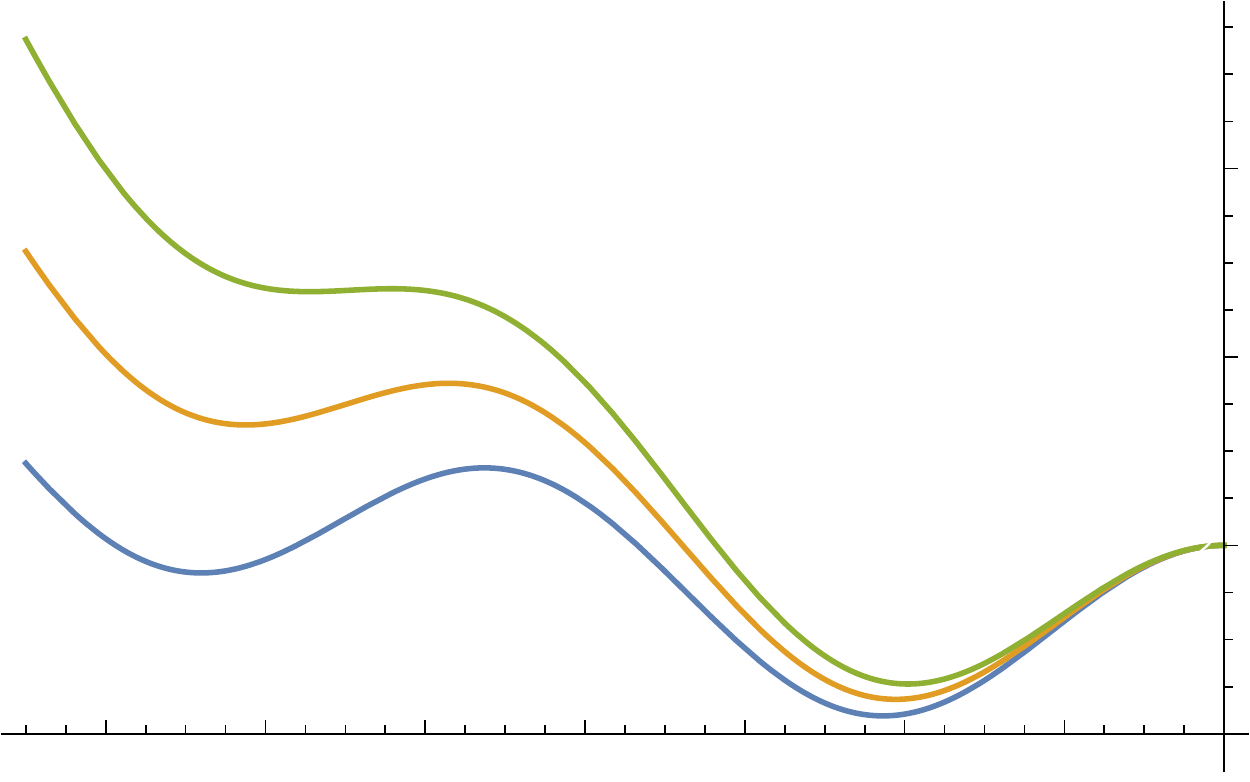} 
\caption{A typical axion monodromy potential as a function of the inflaton field, $V(a)$. In the first figure, we consider the non-thermally 
corrected potential while in the second figure we show the relevant corrections from thermal field theory to the last false minima, close to the reheating 
epoch, triggering 
an efficient phase transition when the thermal corrections are comparable with the local potential curvature. In this plot, we compared thermal corrections in the range $T/(10^{8}\, {\rm GeV})=(0,0.5,1\, {\rm GeV})$,within the illustrative simplified case that the inflaton coupling with fermion species is equal to one. }
\label{f:AM1}
\end{center}
\end{figure}

In this model, the inflaton is an axion-like field that cannot be recast from the solution of the strong CP problem, but is largely discussed in the string 
phenomenology literature \cite{Silverstein:2008sg,McAllister:2008hb,Hebecker:2016vbl,Marchesano:2014mla,Blumenhagen:2014gta,
Hebecker:2014kva,Kobayashi:2015aaa}. Such a scenario is particularly interesting since it is related to a large production of primordial B-modes in the Cosmic Microwave Background radiation. This amounts to a large $r$-parameter, characterizing the ratio tensor over scalar perturbations, which renders 
the model falsifiable in the next future \cite{Silverstein:2008sg}. 

Here, we will show that such a model may be tested also in GWs interferometers, since it is connected to a trans-electroweak phase transition. The scenario represents a particularly important example where our model-independent analysis discussed above can be applied. The axion-monodromy potential has a pattern of periodic local false minima, with the global minima at the bottom of it (see Fig.~3). 

Finally, we emphasize that in the case of warm inflationary scenarios, where thermal corrections are higher than the Hubble scale, the gravitational waves signal may be even stronger, as suggested by Ref. \cite{Bastero-Gil:2016qru}. 

\begin{figure}[!h]
\begin{center}
\includegraphics[width=0.8\columnwidth]{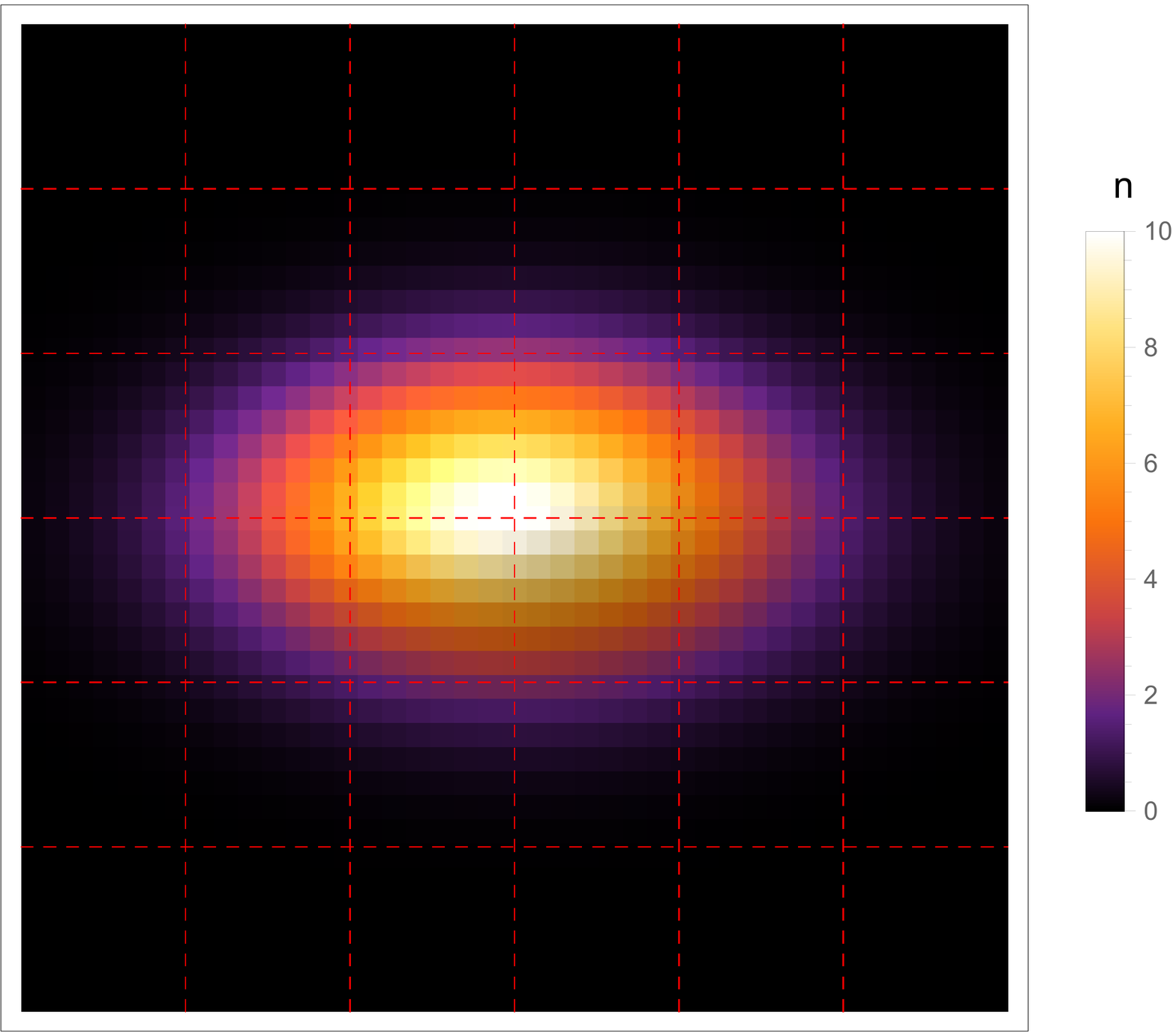} 
\caption{The number distribution of FOPT points as a function of $(x=\beta,y=\alpha)$ axes is displayed, with a corresponding legend. 
The red dotted squares correspond to cell units $3$ (x-axis) and $0.3$ (y-axis) of departures from the number accumulation point.}
\label{f:AM1}
\end{center}
\end{figure}

The inflaton rolls down to the bottom of the potential through a series of tunneling transitions from the local false minima. The last tunneling from the last false 
minimum to the true minimum receives relevant thermal corrections, since it is closest to the reheating process. This can trigger a FOPT
at energies around the typical reheating scale $T_{R}\sim 10^{7}\div 10^{9}\, {\rm GeV}$  \cite{Blumenhagen:2014gta} (see Fig.~3). Thermal radiative 
corrections to the axion-like field are provided not only from the axion self-interactions but also from typical couplings of the axion to the SM fields, 
including 
$$
c_{1}a \bar{f} \gamma_{5} f\, \qquad {\rm and} \qquad  c_{2}a F_{\mu\nu}\tilde{F}^{\mu\nu}\,.
$$ 
Here $f$ are the SM fermions (quarks and leptons) and $F$ are the field 
strengths of SM gauge bosons (electroweak and strong gauge bosons). The coupling constants are not necessarily related to the axion mass and 
the $f$-scale, like for QCD axions, and in principle can be just free parameters of the model. Using the standard DESY corrections, including thermal field theory 
one-loop corrections techniques (see {\it e.g.} Ref.~\cite{Grojan} for applications to electroweak phase transitions) adapted to the axion model case, 
several FOPT points, testable at the ET interferometer, can be found in a large part of the parameter space. 

In Fig.~4, we show the number distribution of FOPT events testable at the ET as a function of the $\alpha$ (x-axis) and the $\beta$ (y-axis) parameters. 
The scan for such events was realized considering the following conditions: axion self-interaction scales as $\Lambda^{4}/f^{2}m^{2}\simeq 50$, $\bar{\theta}=0$, 
$m\simeq 10^{-6}M_{Pl}$; $T_{\rm Reheating}=0.5\times 10^{9}$, and couplings $c_{1f}=0.1\div 1$, $c_{2}\ll c_{1f}$, which correspond to the case of the axion 
strongly coupled to the SM fermions, while all the others couplings with the gauge bosons are negligible. In principle, the axion may be coupled to the Higgs boson, 
but we do not consider this case, since this would turn this scenario into a technically much more complicated one. We found that most of the events are picked around the point $\beta=12.5$ and $\alpha=0.51$, while far from it the viable points disappear. The points found in Fig.~4 correspond to the non-runaway bubbles case. Let us remark that our parameter space scan probably would not include all possible interesting FOPTs for technical reasons but still represents well 
the whole picture and a potential for future experimental analysis of the considering scenario. It would be interesting in the future to further explore even more viable parameter-space islands, which correspond to testable FOPTs. 

\vspace{0.1cm}

{\it Conclusions and remarks.---} We explored the possibility of testing trans-electroweak FOPTs in the early Universe, deploying radio astronomy constraints. The FOPT may occur around $10^{6}\div 10^{8}\, {\rm GeV}$ scales, and generate a characteristic stochastic potentially observable GW background. Depending on the specific subclasses of cases considered, aLIGO and the ET can measure the frequencies around $1\div 100\, {\rm Hz}$, with the possibility of probing FOPT temperature at energies much higher than the electroweak scale, in a range around $10^{8}\, {\rm GeV}$ or so. We remark that our results imply the exciting possibility to test very high energy mechanisms, including inflaton FOPTs, as well as the GUTs, baryogenesis and PBH production models. As an important example, we have made new predictions for the trans-electroweak phase transitions in the so-called axion monodromy inflation and demonstrated its potential for testability at future GW measurements. Our main conclusion is that the new data from aLIGO and the ET will provide a crucial test for FOPTs lying at very high energy scales. This may provide a new important frontier for very early Universe physics. 

\vspace{0.1cm}

{\it Acknowledgments.---} A.A. and A.M. acknowledge support by the NSFC, through the grant No. 11875113, the Shanghai Municipality, through the grant No. KBH1512299, and by Fudan University, through the grant No. JJH1512105. R.P.~was partially supported by the Swedish Research Council, contract numbers 621-2013-428 and 2016-05996, by CONICYT grant MEC80170112 (Chile), as well as by the Ministry of Education, Youth and Sports of the Czech Republic project LT17018.

\end{document}